# Technical performance and interpretation of physical experiment in problems of cell biology

Dalibor Štys[1]

**Abstract.**

The lecture summarises main results of my team over last five years in the field of technical experiment design and interpretation of results of experiments for cell biology. I introduce the theoretical concept of the experiment, based mainly on ideqas of stochastic systems theory, and confront it with general ideas of systems theory. In the next part I introduce available experiments and discuss their information content. Namely, I show that light microscopy may be designed to give resolution comparable to that of electron microscopy and that may be used for experiments using living cells. I show avenues to objective analysis of cell behavior observation. I propose new microscope design, which shall combine advantages of all methods, and steps to be taken to build a model of living cells with predictive power for practical use.

## 1 Introduction

Biology in since around 1955 has been dominated by molecular techniques. It has been the success of cracking the genetic code [1] and namely the technical developments in gene sequencing [2] which lead to imaginations such that sequencing of individual genome will answer all questions of life. In medical practice, for example, the identification of living system state with its genome is still dominating. It may be demonstrated on the basically correct idea of personalised therapy, which reflect the fact that some patients react positively to certain therapy while for others it may have fatal consequence. Knowing the individual state of the patient may help to predict the outcome of use of certain drug. However, the term personal state is almost unanimously identified with personal genome. Despite to the fact that we know from everyday practice that for living system is much more than for anything else on Earth true that system is much more than collection of its parts [3]. Sticking to the systems approach fact causes opposite reaction, equally incorrect. This is partial or rather large ignorance of conflicting laboratory facts by medical practicioners. It is the absence of experimentally testable system model of the living organism which is in the core of this misunderstanding. And, as I would like to demonstrate in this paper, we are technically approaching the possibility to build such a model and test it.

Since late 1990´ the new field of systems biology arose [4]. It has been evolving around the models of metabolic control developed in 1970´ [5] and evolved in parallel with theory of non-linear dynamics and complex systems [6]. From the historical interest it should be mentioned that the first approximative model of life-like structure in chemical mixture, the chemical clock, was presented in 1968 in Prague by Anatolij Zhabotinsky [7]. The later example illustrates the problem which arises when one tries to develop models of macroscopically simple behavior, such as chemical clock or cell cycle, on the basis of microscopic observation. The actual mechanism of chemical reaction is quite complex [8] but may be qualitatively, but much extent also quantitatively, modeled by relatively simple agent-based model [9]. The chemical clock agent model including just three probabilities of state change and multiple numbers of available states. In biological applications, the cell state is the macroscopic behavior and the metabolic and signal cycles are the origin of probabilities of its change.

In this lecture I would like to discuss methods by which we analysed current state of the development of biological model from the point of view of information content. From the point of view of biology, our reasoning is based on the simplest medically relevant example. In this way we are always confronted with everyday medical experimental reality. We believe that by this approach we keep perspectives for correct interpretation of biological experiment in general, applicable for bioreactor control, medical diagnostics and all other biology applications.

---

[1] Institute of Physical Biology, University of South Bohemia, Zámek 136, 373 33 Nové Hrady.

# 2 Systems biology, biological experiment and mathematical models in biology

## 2.1 Models, modeling and experimental reality

It is generally assumed that mathematical modeling helps us to understand internal nature and dynamics of observed processes and to formulate predictions about their future development [10]. The term model means different things in different communities, here will be discussed that which prevails over disciplines ranging from experimental biology, through chemistry up to experimental physics. The question frequently asked by cognitive scientists is why is mathematics so useful. According to Wigner [11] most theories owe their success to mathematical descriptions devised without any idea that the will eventually be applied. Over the 19$^{th}$ and 20$^{th}$ century it prevailed the opinion, that model is an abstract representation of objects and processes that explains features of these object and processes. Or, in another words, that it reflects the underlying phenomena.

The realistic view of model has been always matter of discussion. The most striking example is quantum mechanics, which from its beginning [12] has been considered by some of its key protagonists only a useful model. Now, there have been proposed a few models which explain quantum mechanics equally successfully without introduction of many of the specific axioms [13, 14]. This type of discussion occurs in biology only *in principle* [4]. In reality of the field of systems biology there are two different fields, the molecular modelers, which reduce biological models to chemical models, and biological experimentalists which use qualitative descriptions.

Perhaps the most visionary definition of the model in experimental science we may find in the work of Bohemian scientist Georg de Buquoy (1781-1851) [15] who proposes the method of "parallelisierung" by words: "The unavoidable utilization of mathematics comes mainly from dominant spatial spread of the inorganic (better sub-organic) *(form of matter)*. But what maters the laws of other phenomena, the qualitative laws of spatial *(mechanical)* phenomena have their analogies in laws of life. Mediating utilization of mathematics is here only allowed. The analogy is the only thing which should be sought here. And the utilization of mathematics at higher opinions about the natural life should not go any further beyond the aspiration to paralelise the laws of mechanics with their derived laws of relations and quantitative *(predictions)* with laws of living nature and its relations. …….. In any use of mathematics for paralellisation one must be aware that the mathematical forms may be used not only as mediating symbols (which is often seen) but that they must be understood in the full spirit of their meaning. In the same way as for the geometer can use the script with full mathematical correctness and not only as meaningless hieroglyphs. " In modern wording the prototype model should be sought in mechanics (but in Buquoy´s, very general, sense) and its consequences should be fully explored.

The meaning of words inorganic is constrained to strictly mechanic phenomena by definition found in [15] "We probably may say, that in each actions in the nature there is certain degree of egoism, prevalent subjectivity in the point of action, with exception of one action, movement through mechanical power in non-living bodies." Which is almost identical to the formulation of the second law of thermodynamics by Caratheódory who says [16] "In any neighbourhood of any state there are states which can not be reached from it by an adiabatic process." Yet, there is a substantial difference between Buquoy and Caratheódory, the generality. What Caratheódory still constraints to phenomenon of heat, Buquoy already expands to chemistry, biology an economy. We can value the modernity of this approach only now, with the eve of systems engineering, econophysics and complexity science. At least from the time of Jaynes [17, 18], it is clear that there is connection between uncertainty of the model – probabilistic behavior of experiment outcome and the "adiabatic unreachability" of Caratheódory [16].

Thus, we have two interconnected ideas. (a) modelling as way to examine some important features of the internal structure of measured system which we can not measure precisely and (b) inherent, thermodynamic or quantum mechanical, uncertainty of the system which can not be overcome. We can deal with these phenomena in an engineering way. That brings about the theory of stochastic systems [19], which was precisioned to great exactness by Žampa [20-22]. In this engineering approach, the stochasticity is strictly attributed to our inability to measure precisely. Any real life experiment lacks technically feasible infinitely small yardsticks and any real measurement is realized with finite time resolution.

For the description of technically constructed systems is this technical explanation sufficient. But it does not seem to be complete, since we do observe spontaneous mixing of two solvents and spontaneous temperature equilibration in two heat baths. In a sense, the principle of quantum mechanics when applied to explanation of thermodynamic phenomena, and when re-phrased in the terminology of stochastic systems theory, says that nature is permanently performing individual experiments and that it has a primitive yardstick. The length of this elementary yardstick is called the Planck constant $\hbar$ and the explanation of the elementary imprecision is called the Heisenberg uncertainty principle [23] which says that

$$\Delta p \Delta x \geq \frac{\hbar}{2} \tag{1}$$

where $\Delta p$ and $\Delta x$ are standard deviations in determination of moment and position respectively. Heisenberg derivation implicitly assumes normal distribution of probabilities. This can not be assumed in many measurement processes, the probability density function must be examined for each experiment, the theory of Zampa [20-22] assumes this quite naturally. Also for the "natures primitive" experimentation this is necessary to consider [24]. We then have to use equation

$$H^{(x)} + H^{(p)} \geq 1 - \ln(2) - \ln\left(\frac{\delta x \delta p}{\hbar}\right) \tag{2}$$

where $H^{(x)}$ and $H^{(p)}$ are Renyi entropies appropriate to measurement of position and moment and $\delta x$ and $\delta p$ corresponding statistical dispersions. In real world there is nothing like purely mechanical process, any observed process combines thermal/entropic and mechanical component. The difference between the textbook statistical thermodynamics and Buquoy´s paralellisierung method lies mainly in the focus to energetics (a scalar variable) and omission of dynamics (equivalent to moment). In other words, Buquoy proposes to examine the whole phase space, standard approach of contemporary thermodynamics is charting the state space.

In cell biology – but also in economy, population ecology or even sociology or behavior of traffic jams, we observe temporarily stable structures. They are often of cyclic character (or have cyclic internal dynamics) and loss of their stability may be described by probabilistic formulas. Analogous phenomena in mechanics may be found in non-linear dynamics [25], according to which the state space may be dissected into sub-spaces belonging to individual limit cycles. Limit cycles are trajectories in the phase space along which the system returns to nearly identical position, a similar sub-set of points in phase space. The trick is that after few orbits, it is not any longer possible to trace back the precise original trajectory, there is a whole set of initial trajectories which give rise to given asymptotically stable cycle. The system leaves the limit cycle with certain probability which is described by typical PDF, probability density function.

In multidimensional systems with a few spatial and a few chemical state variables, the PDF is obviously multidimensional too. However, since (a) in limit cycles the system resides for much longer time than in any unstable state and (b) since our yardstick may be too rough in time and in space and we are not able to distinguish individual trajectory, we may in reality observe only limit cycle and omit other points in phase space which are extremely rare and thus seldom observed. The Boltzmann distribution in thermodynamics is based on essentially the same assumption [26]. The limit cycles naturally emerge. The later term may also be formulated in the way that we may quantify the system behavior only within certain limits. The beauty of this analogy – or parallel - lies in the creation of inherent measure, limits, within we (for example) call the cell cervix tissue cell and outside of them we call it cervic cancer cell. It is the number and identity of limit cycles within the cell state space which make the difference. We may never know all of them and even less probably we shall be able to compute their properties. What is, nevertheless, striking is the fact that the mathematics of the non-linear dynamics may be utilized also for non-axiomatic formulation of quantum mechanics [13, 14] and therefore we may consider that the phenomenon of internal yardstick is a general one, substantiating the qualitative difference between natural phenomena.

My proposition given here is that each process has its internal non-linear dynamics which gives rise to its emergent behavior. This defines not only observable states and probability density function of transitions between them, but also, and mainly, the appropriate yardstick in mechanical and chemical space by which the state of the system should be measured. This includes also minimum time of observation or, in another words, minimal time interval for which the system must be measured to ensure appropriate observation of system behavior.

In any case, the internal dynamics will dictate the appropriate probability density – or state –function, the norm by which limits the precision by which it may be measured and appropriate Renyi entropy [24] by which it should be evaluated.

## 2.2 Experiment in cell biology – statics and dynamics

Whatever it may be tempting to expand on bird flock formation or tree canopy, I constrain my reasoning to cell biology experiment. And among it, to experiments performed in my laboratory, light microscopy of living cells, and proteomics and metabolomics performed by liquid chromatography – mass spectrometry (LC-MS) of cell extracts. As reasoned in part 2.1, we must consider two limits of observability of the system. The technical one, given by the construction of our microscope and the LC-MS chromatograph and that given by the biological system, by properties emerging from its dynamics. The text will be focused on discussion of results of cell development dynamics observed by optical microscopy, namely since the discussed phenomena may be best observed using this technique.

We may summarise the problem of observation in the frame of general systems theory by Žampa [20-22]. The basic experiment in cell biology, observed by time-lapse microscopy, looks typically as shown at Supplementary video 1. We demonstrate typical monolayer behavior on the example of He-La cells, standard human cell line used in basic cell monolyer experiment. He-La cells are human cervic cancer cells cultivated since the year 1955. Cells divide and integrate into a monolayer. Basic classification of cell behavior is to separate cells into mitotic cells and cells in the interphase (figure 1).

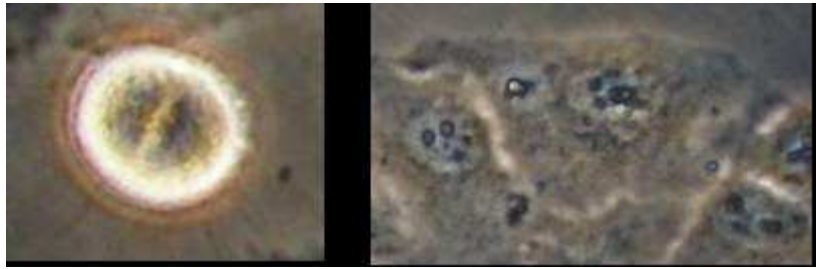

**Figure 1** Mitotic cell (left) and cell in interphase (right). For further discussion it should be noted that cells in interphase have fuzzy borders in the phase contrast image and have also weakly contrasting borders between cell interior and background.

By detailed inspection one may also observe physical contacts between cells, so called pseudopodia (figure 2) and find relations between mother and daughter cells (figure 3).

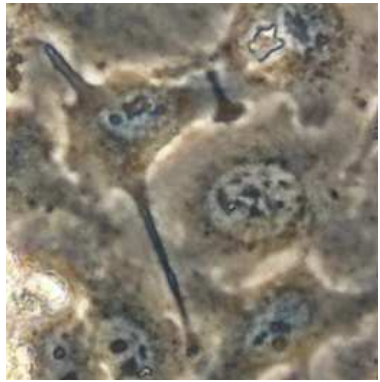

**Figure 2.** Pseudopodium by which non-neighbouring cells exchange matter. The purpose of this matter exchange is probably informational. Similar transfer of matter may be observed across the borders of neigbouring cells in interphase.

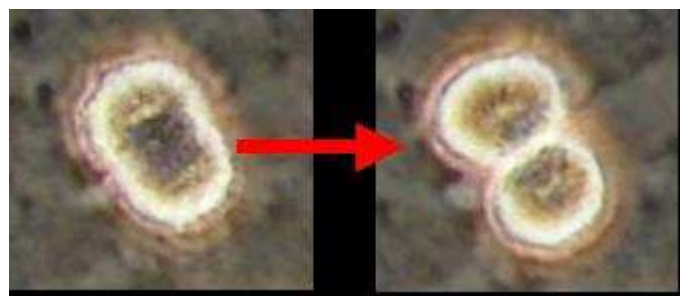

**Figure 3**. Formation of two daughter cells under standard cell division.

Cells are clearly non-linear dynamic systems. It is also rather clear that we shall never be able to observe concentrations of all compounds in each of the organelles with infinite time and space resolution in vivo. Yet, we are interested in the model of the biological system with reasonable predictive power useful, for example, in medical diagnostics or bioreactor control. And this model should not be dependent on the performance of the microscope or provision of the chromatographic experiment. In this chapter I wish to outline the biological model.

According to Žampa [20-22], in any real life experiment we observe attributes of the system $a_i \in A$ reflected in a set of adequate variables $v_i \in V$. The set of attributes $A$ and the set of variables $V$ are homomorphous, indexes $i \in I$ are parts of adequate index set. We also define ordered set $T$ of time instants $t_j \in T$ for which it holds that $t_0 < t_1 < .... < t_{max}$ - which covers the whole measurement interval. Set

$$D = T \times I \qquad (4)$$

is the definition set of the system. We may then define mapping

$$z: D \to V \qquad (5)$$

which we call system trajectory.

For our concrete example (supplementary video 1) the set $T$ is sequence of times from the beginning of the experiments up to its end, 300 000 s. With uniform time interval 60s. In most cases we observe small differences of individual objects in space and in time. Let us call this situation continuous cell monolayer development. At certain time instants we observe large changes in the scene, thus within this time interval obviously happened several changes equivalent to those observed during the continuous development. These changes were not captured due to defined time resolution. The appropriate model of an experiment has to consider this.

The system trajectory is composed of trajectory elements $(C_{k,l}, D_{k,l})$. As explained at figure 4, the sub-set $D_{k,l} \in D$ is a set variables values observed at evaluated time of measurement and the sub-set $C_{k,l} \in D$ is a set of all variables observed at times $t_h < t_k$ necessary to predict unanimously the future behavior of the system.

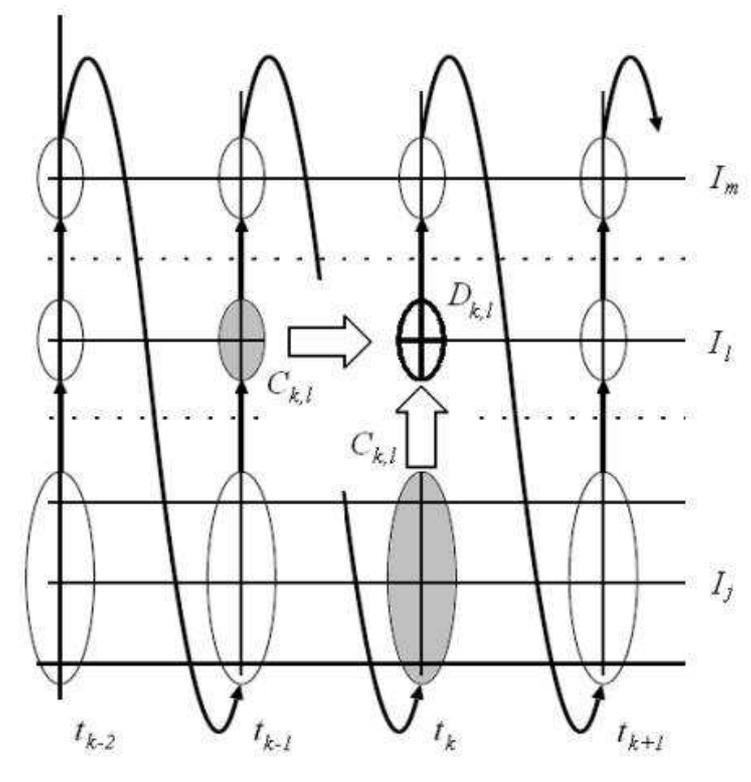

**Figure 4.** Complete immediate cause. For the sub-set $D_{k,l} \subset D$ we seek all variable values which are needed fort prediction of future behavior of the system. To define state of the system in the same sense as in mechanics or thermodynamics, future values of state variables must not depend on path by which the current state was reached. To ensure that, we need to take into account so long part of the system history that variable values at any preceeding time are irrelevant for future prediction. The main considerable problem is that of discrete time intervals. First the system develops between measurement time intervals and there are variable values which we either do not measure, or we measure them in some aggregate form. There is averaging included in the performance of the measuring device and there is a minimum time, determined by the nature of the system itself, for which the measurement has to be performed. This is in the figure represented by the bottom arrow. It may also be possible that the necessary measurement time is longer than one time interval. Or, even more strikingly, we need sampling over several time intervals (i.e. in oscillatory behavior). Then we need for determination of system state also measurements from preceeding state intervals. At that from as many preceding time intervals, as needed for the probability behavior to be completely Markovian, independent from history. Fortunately, in non-linear system the history is in most cases forgotten after certain period of time.

To find the trajectory element, we seek ordered subset $(C_{k,l}, D_{k,l})$, in which both $C_{k,l}$ and $D_{k,l}$ are part of the system definition set $D = T \times I$. D is Cartesian (= vaguely saying ordered) product of set of all time instants T and the set of all attribute indexes I. $C_{k,l}$ consists of all necessary time instants and attribute indexes necessary to calculate unanimously the $D_{k,l}$ by the model of system trajectory. System trajectory is mapping $z : D \to V$ where V is set of all variables. System trajectory consists of trajectory elements $(C_{k,l}, D_{k,l})$. The trajectory element thus has its own, inherent, timespan, which may be shorter than difference between two time instants, as well as longer.

Although apparently complex, the analysis of the biological system based on the approach of generalised stochastic systems theory is essentially equivalent to biological intuition. Natural elements of the trajectory are elements of the cell cycle which have different names for different organisms or the cell cycle itself. While the elements of the cell cycle as system states are not questioned, the identification of cell cycle as cell state requires experimental verification, as will be shown later. In the case of He-La cells the elementary states in the cell cycle are called mitosis and interphase. Naturally are observed events like cell death, cell fusion, anomalous cell division etc. The reality is that the manual analysis of the 3500 to 5000 images allows the operator to notify only mitosis and interphase and the cellular neighbourhood of observed cells.

Figure 5 shows a series of datasets in the course of manual analysis and example of corresponding sub-graphs. The sub-graph showing fate of the cell followed from the point of view of the individual cell as system element (Fig. 5b) is a reflection of the model according to which cells communicate only by direct long-term physical contacts. The cell fate itself and any change in number of its neighbours are noted. The sub-graph showing cell fate from the point of view of the whole system (Figure 5c) is an expanded version based on the assumption that any change in any cell is notified (i.e. through signal molecules) by all the cells.

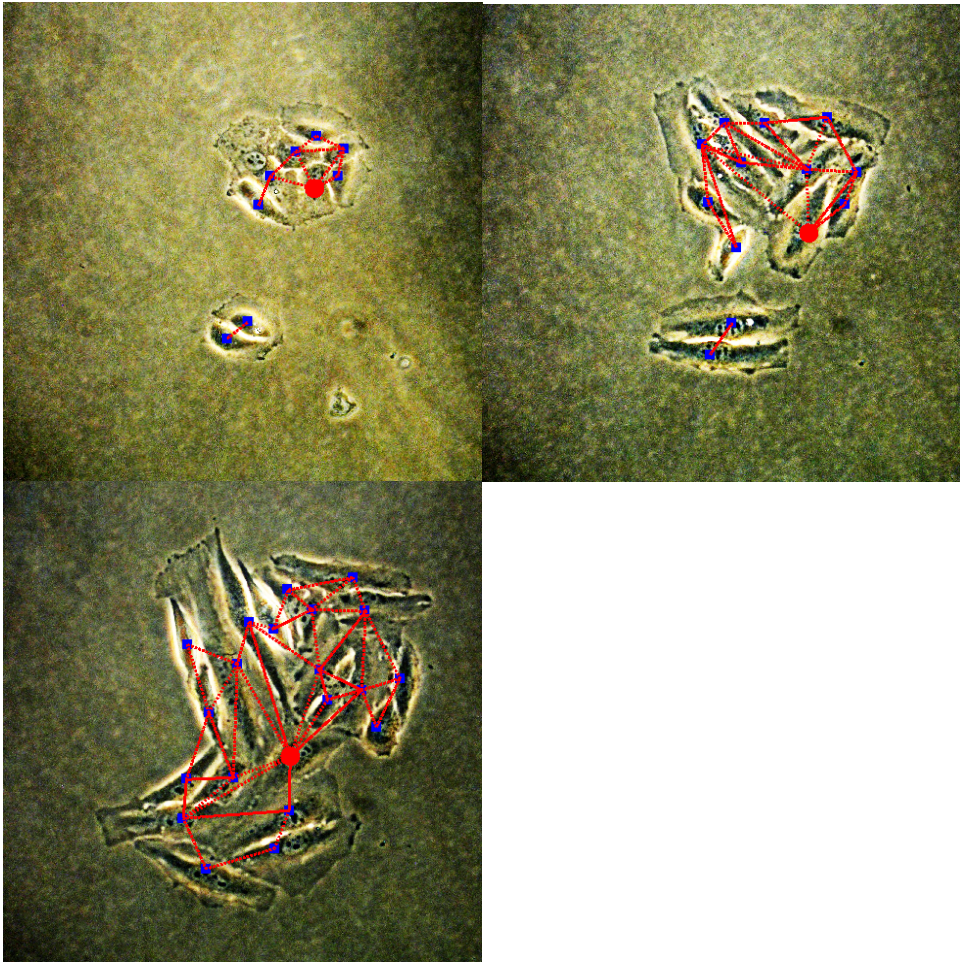

**Figure 5a**. Cell monolayer culture image No. 2 (1 min), 1000 (999 min) and 2500 (2499 min) of the time-lapse experiment. The cell discussed in subsequent graphs is denoted by the red dot.

From data obtained in this way, one may produce plot of length of these elementary trajectories versus experiment time (Figure 5b and 5c). The initial hypothesis was that there is certain "bounded asynchrony" in the cell development, as was proposed by Fisher et al. [27]. This term basically means that there is an inherent time of the cell cycle which is shortened or elongated by signals received from other cells. Since there was no guidance how to construct the model from the data, we decided to follow strictly the receipt of generalised stochastic systems theory.

We assume that the system may be separated into sub-systems $\pi_r$ which (a) in absence of the rest of the system will evolve independently and (b) can not be further separated into sub-systems. The set of all such elementary sub-systems is called system universum $\Pi$, and it holds that

$$\pi_r \in \Pi \quad r \in N \tag{4}$$

$\pi_r$ is the sub-system and $N$ is a sub-system index set. The sub-systems interact by information bonds

$$c_{k,l,q,s} \in \left(C_{k,l} \big| D_{k,l}\right) \tag{5}$$

where indexes $q, s \in R$ indicate the identity of sub-systems. The information bond is generally expected to be directional.

Second major assumption of the stochastic systems theory is the stochastic causality. Each tuple $\left(C_{k,l}, D_{k,l}\right)$ may be called causal relation and the mapping

$$\Xi : D \to P(D) \tag{6}$$

where $P(D)$ is the potency of the set, set of all sub-sets which give rise to different trajectory elements, is called causality. It basically says that the whole definition set $D$ is segregated into sub-sets between which there are non-uniform transitions. The causal system $KS$ then may be described by:

$$KS = (T, V, \Xi) \tag{7}$$

The purely mechanical system observed in continuous time would be deterministic. For all real systems the trajectory element must be attributed by its probability density which reflects the measurement inprecision, i.e

$$p_{k,l} = p(z|C_{k,l}, z|D_{k,l}) \tag{8}$$

and

$$\mathrm{P} = \{p_{k,l}\} \tag{9}$$

Is set of all available probability densities in the system. Finally, the stochastic causal system is defined as

$$SKS = (T, V, \Xi, \mathrm{P}) \tag{10}$$

The sub-system $\pi_r$ may be considered primitive stochastic causal system which in absence of information bonds will evolve independently

$$\pi_r = (T, V_r, \Xi_r, \mathrm{P}_r) \tag{10}$$

Quite naturally, the biologists would assume cells to be sub-systems and observed transfer of matter between cells to be the information bond. Independent evolution of cells will occur in "unbounded" uniform cell cycles. The information bond may also be general, the release of a signal molecule to the surrounding space for example. In the case of tissue culture or cell monolayer we assume that all sub-systems are equal. We have observed that overhelming number of mater transfers occurs through the wall of two neighbouring cells, only very few through pseudopodia of non-neighbouring cells.

On the basis of this reasoning we created a computer program *Expertomica Cells* [28] in which we analysed the fate of individual cells and evolution of cell neighbourhood. The probability of cell division was approximated by the time between one and the following mitosis. Graphical representation of such results is always a matter of a simplified model, we present here two models, local and global. In the local model (figure 5b) we assumed that information bonds are only those which are directly connecting the neighbouring cells (i.e. through observable channels between cells) and that probability of formation of these channels is equal between each two cells. In the global model (figure 5c) we expected that main information bond reflects the occupation of the space by cells. In another words,, information bonds were assumed to be long-range,. i.e mediated by communication metabolites released by cells to the medium. And we expected that introduction of an information bond changes the probability of cell transition, shortens or elongates the time interval between mitoses.

However, when we evaluated these results, we have not observed any intelligible dependency of the length of the cell cycle on experiment time (Figure 6), on number of neighbours, total number of cells or any other parameter which may be derived from the analysis.

Perhaps this result should not really be surprising. By detailed inspection one may find that in majority of articles which try to diagnose tissue cultures from times of cell division, after all the conclusions are made on the basis of modelled data. Thus, the simplified bounded asynchrony model has no correspondence in real data.

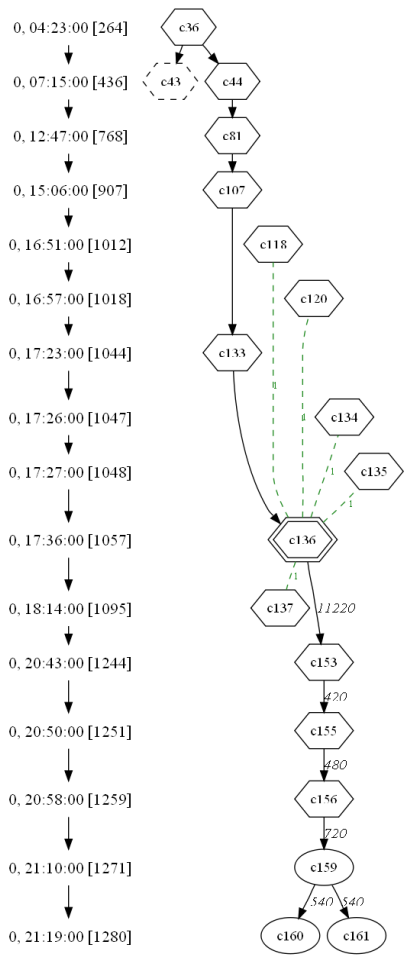

**Figure 5b. Fate of the cell followed from the point of view of the individual cell as system element.** Solid line represents one single cell fate, dashed lines represent cell neighbourhood. Complete information is depicted only for the marked cell.

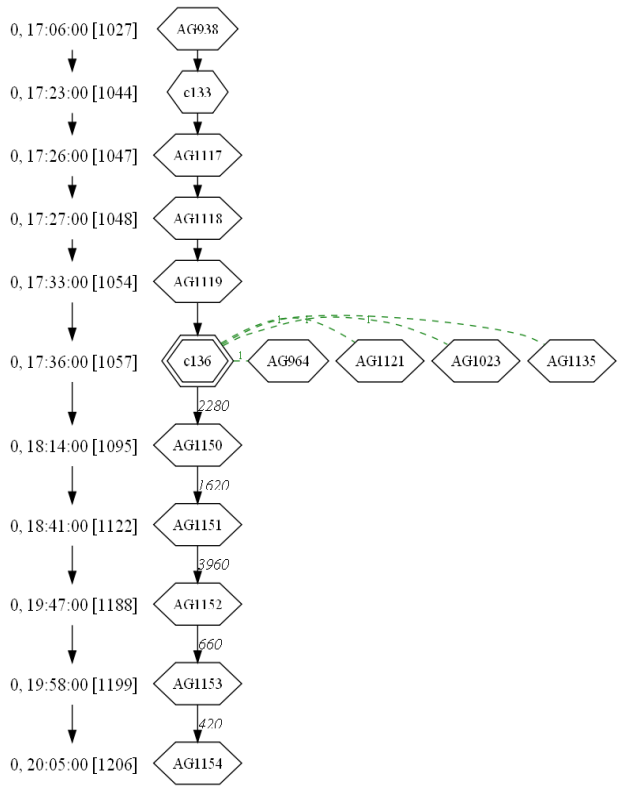

**Figure 5c. Cell fate from the point of view of the whole system.** Cell numbering beginning with AG denotes automatically generated. Times of these AG cells represent any change in any cell in a particular neighbourhood network. By neighbourhood network we understan a cell cluster, all cells who are in contact either directly or mediated by other cell or cells.

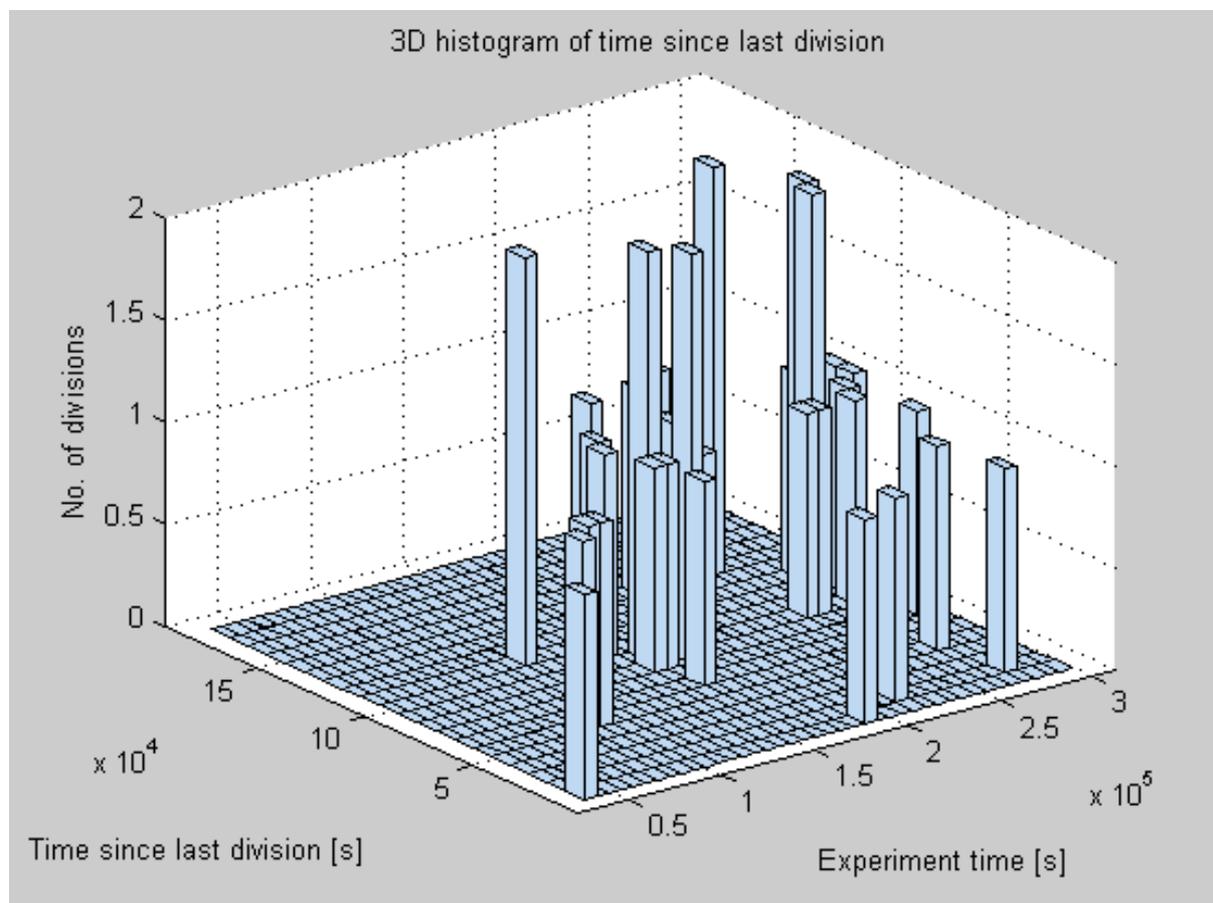

**Figure 6. Dependency of cell cycle time on experiment time**. Originally certain degree of synchronicity and uniformity of cell cycles was expected. However, this was not observed.

## 2.3 Microscopic experiment in biology and its information analysis

The above mentioned failure of the manual analysis of time-lapse observations of development of cell monolayer shows that simple observation of the cell shape is not a good indication of the cell state. The scientific literature, however, leaves us with empty hands when we search for other indices. Certainly we may find articles in which the time-lapse microscopy was utilised, in recent times they are mainly focused on fluorescence microscopy [29], where it is possible to rather clearly identify the signal with concrete molecule. We did not want to follow this path for many reasons. The most important for us was the preservation of comprehensiveness – one molecule usually represents only partly and non-specifically one process in the cell [30]. Equally important was also our intention to work with unlabelled and chemically unmodified cells. Such cells may be derived from real patients and stay in the state close to the natural one. Here, I will outline only those features of light microscopy in biology which are important for representation of our results.

**Resolution and discriminateability in optical microscopy** [31] has been an issue for at least two centuries [32, 33]. Basic ideas are still arising from the work of Airy [32] who rather correctly devised a model, based on light interference, which explained why an object is imaged in a form of concentric rings. His theory was based on the model in which image is formed by light interference. His idea was the first on the avenue towards modern formulation of the point spread function (PSF) [34]. The microscopic community is somewhat biased by the Abbe formula [33] which is little more than great simplification of the original Airy idea. For Airy, the resolution of the microscope is defined as the distance between centres of two diffraction images of a point (concentric rings) in the case when exactly the valleys between the zeroth order and first order ring for the two objects exactly merge. Today, this idea is extensively applied in fluorescence microscopy where we may ensure that we deal with separate point sources of light.

In standard light microscopy the image is formed mainly by diffraction of light at edges of object with different refractivity index. Some assumptions of Airy can not be guaranteed. Namely we should ask question what happens when the diffracting objects are side by side and do not have ball shape. When the two point spread functions (in this case we should rather use the term object spread functions) meet, we certainly observe intensity drop. But where exactly this apparent border resides is given by the course of points spread functions in the particular geometry. We propose the term discriminateability for our ability to see two objects without being able to exactly describe their shapes.

Alternative formulation of principles of microscope resolution has been increasingly based on uncertainty principle in quantum mechanics. The interest has been risen by the observation of objects of the size of 10-20 nanometer by videoenhanced microscopy [35]. Mizushima [36] explained this observation by principles of quantum mechanics. He practically neglected previous derivations and using the equation (1) he came to formula

$$\Delta x \geq \frac{\hbar}{2\Delta p_x} \qquad (11)$$

Where $\Delta p_x$ is the moment uncertainty in direction perpendicular to cell axis. The moment of light is proportional to $\sqrt{N}$ where $N$ is number of independent photons arriving to the detector at the time of measurement, the resolution at high light intensity is virtually infinite. This derivation also became more or less part of the common knowledge, but it has the same problem as many similar derivations based on quantum mechanical ideas, the improper treatment from the information theory point of view (see eq. 2).

The ultimate answer is given by the experiment. Our approach is based on computation of the information contribution of individual points to the information brought by the image. The approach was described in [30]. In brief, we quantify occurrences of individual intensities in the image with and without examined point. Probability is approximated by division of occurrences by total number of points. Resulting numbers are used for calculation of image entropy. Difference between entropies calculated from probability distributions with and without examined point is the information contribution of the point to the image. For evaluation of the image information we use Renyi entropy [37]

$$H = \frac{1}{1-\alpha} \log_2 \sum_1^n p_i^\alpha \qquad (12)$$

We use multiple $\alpha$ parameters and compare the results. This is in full agreement with the theoretical assumption that the information space of the microscope has multifractal character [30].

In technical terms, the microscope transmits a whole section of space which is defined by the size of the scene and the depth of focus. It includes objects lying exactly in the focal plane, below and above it, in the centre of the scene and on its edges. At each of the positions with respect to the focal plane and of the lense axis, the point object is transmitted as different, generally non-homogeneous, set of points. The resulting image is a mixture of these responses. In case that a space has non-homogeneous fractal character, its dimensionality [38] is generally given by formula

$$D_\alpha = \lim_{d \to 0} \frac{H_\alpha}{\log_2 d} = \frac{1}{1-\alpha} \lim_{d \to 0} \frac{\log_2 \sum_1^n p_i^\alpha}{\log_2 d} \qquad (13)$$

Where $d$ is the yardstick length, the length of measure which we use to cover the given space. Thus, there is intimate correspondence between the Rényi entropy and the fractal character of the space. If there are more different fractal measures, there are also more different appropriate Renyi entropies. The multifractality in the image space must not necessary come from the information channel – the microscope optics. It may also originate from self-similar objects inside the observed specimen, for example the living cell.

**Technical limits of present day optical microscopy** may be demonstrated practically on concrete images taken by the videoenhanced microscopy. At the figure 7a we see section of the image of the yeast cell in original brightfield version and in version transformed using Renyi entropy with $\alpha = 0.1$. At the figure 7b there is the same image with $\alpha = 2.0$.

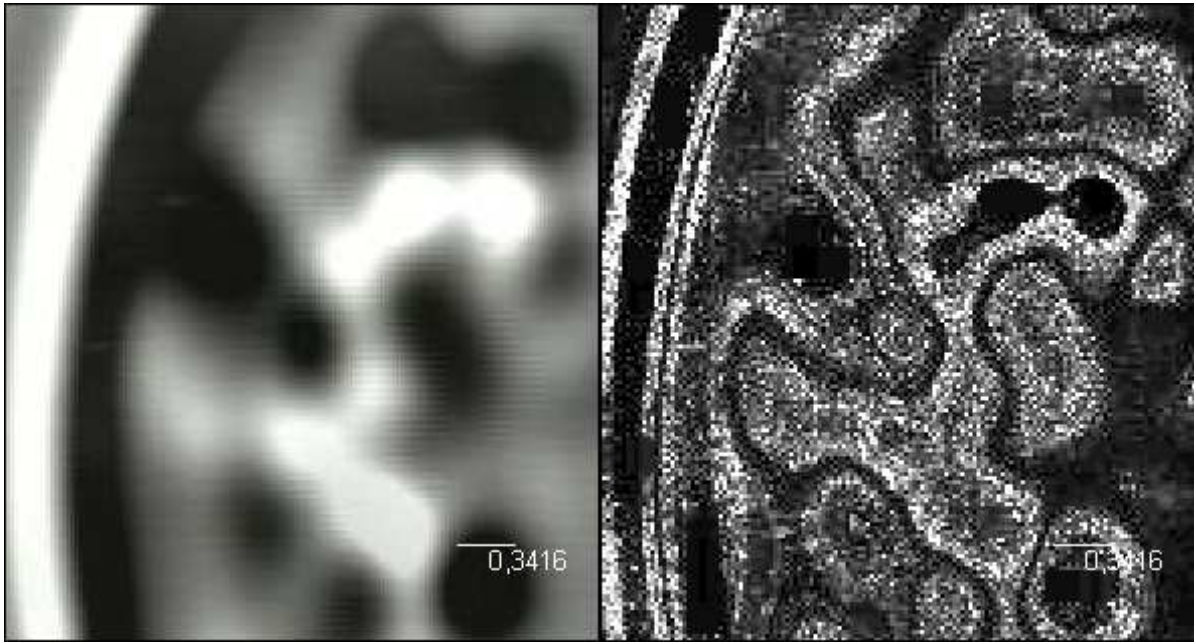

**Figure 7a. Image of a section of yeast cell** interior, cell wall and of the background medium in the original brightfield micrograph (left) and in the Renyi entropy measured image information representation. The Renyi entropy coefficient $\alpha = 0,1$. Optical magnification was 100 x 1.6 x 1.6 where 100x magnification was achieved by the immersion objective lense and the rest of the magnification by combination of ocular lenses.

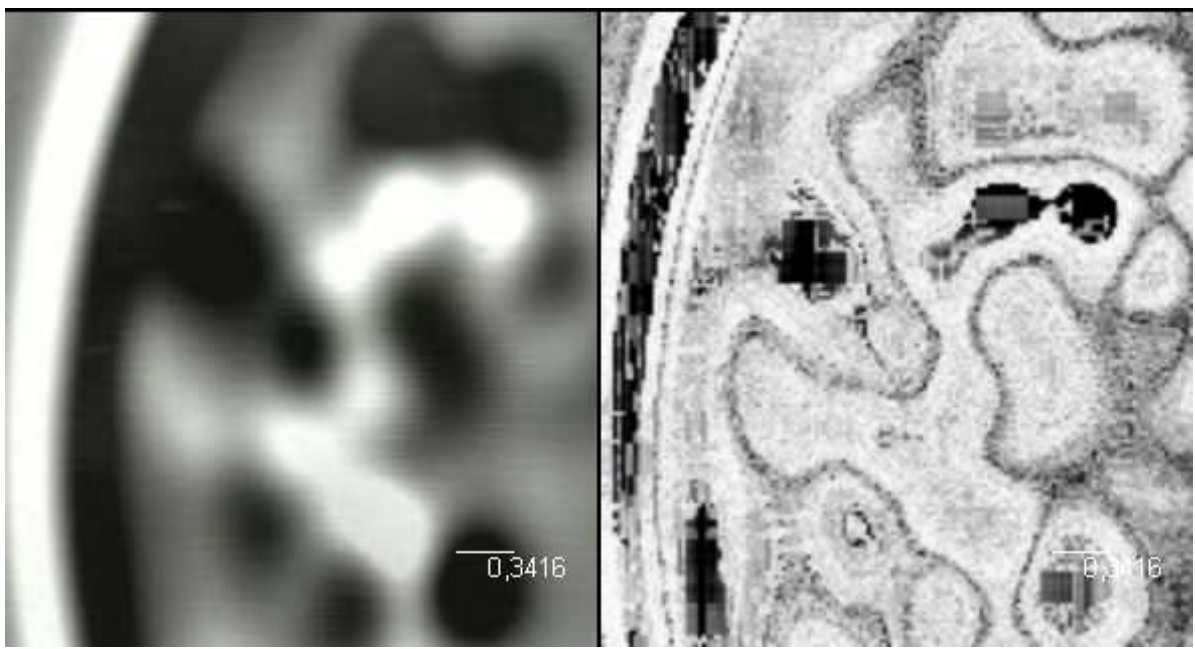

**Figure 7b Image of a section of yeast cell** interior, cell wall and of the background medium in the original brightfield micrograph (left) and in the Renyi entropy measured image information representation. The Renyi entropy coefficient $\alpha = 2,0$. Other measurement conditions were the same as in fig. 7a.

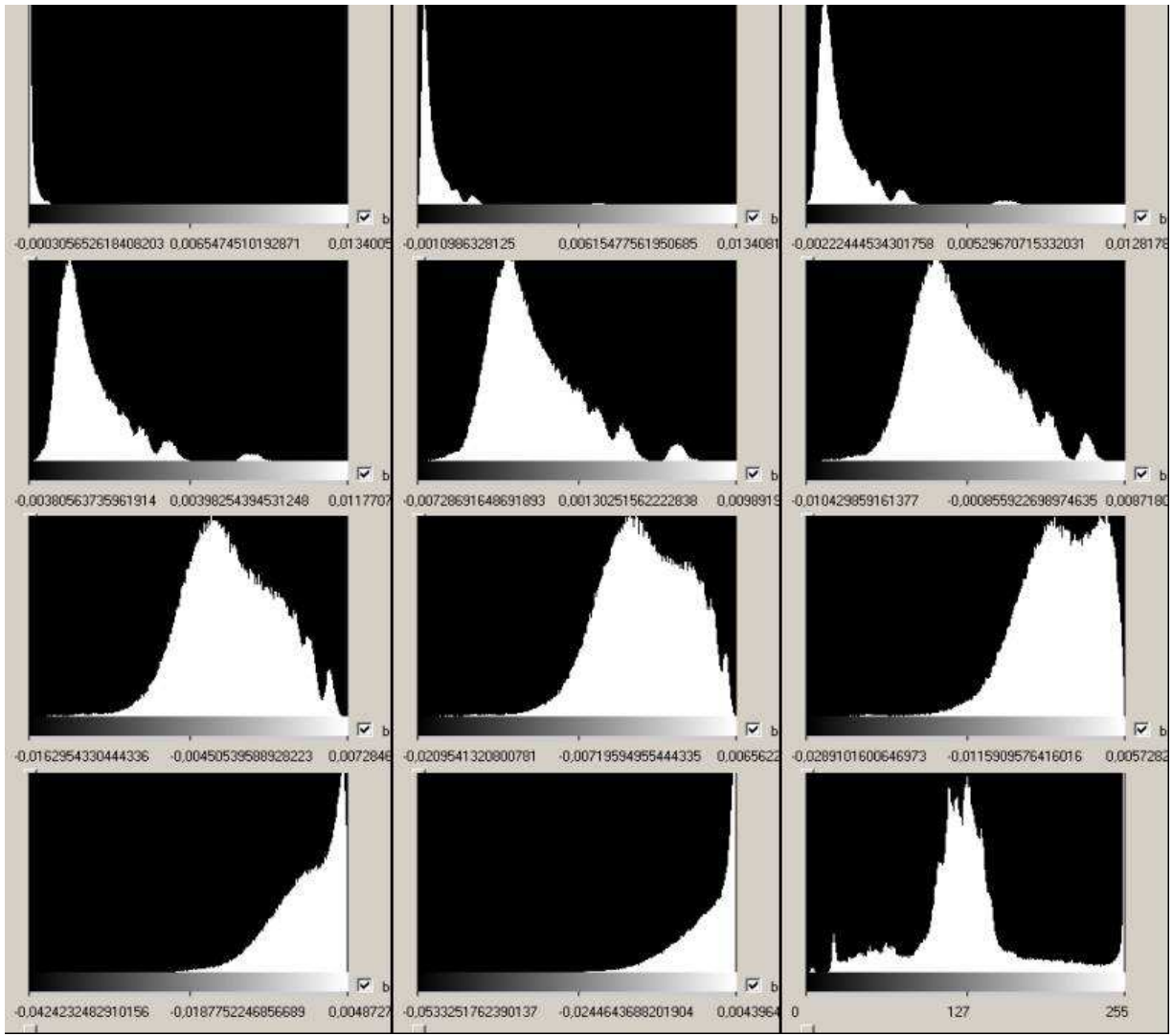

**Figure 8 Image entropy spectra** of the image of the yeast cell. From upper left to lower right the $\alpha$ values are 0.1, 0.3, 0.5, 0.7, 1.0, 1.2, 1.5, 1.7, 2.0, 2.5, 3.0. The last image is histogram of original intensities.

Perhaps more interesting are, however, distributions of entropy spectra. It is clear that we may observe entropy spectrum but it was not obvious whether we shall see one uniform distribution or several pekas. Each of the peaks in the spectrum represents certain type of features which have similar probability and these features may be extracted from the image. For illustration we show on the figure 8 the same image as in the Figure 9a depicting points contributing to one selected entropy maximum for Rényi entropy with $\alpha = 0.5$. The corresponding histogram is shown at figure 9b. Detailed examination of images in different entropy spectrum shows that some of the features are represented only in certain parts of the entropy spectrum. The best impression of the meaning of a section is observation of these points in dynamic cell behavior (see supplementary movies 2 and 3). In the particular case discussed in this section, we may observe movement of certain representative points in the cell. Unfortunately, however, the exact value of the image entropy varies with technical setup, namely since the extent of cell representation varies with its position relative to the depth of focus.

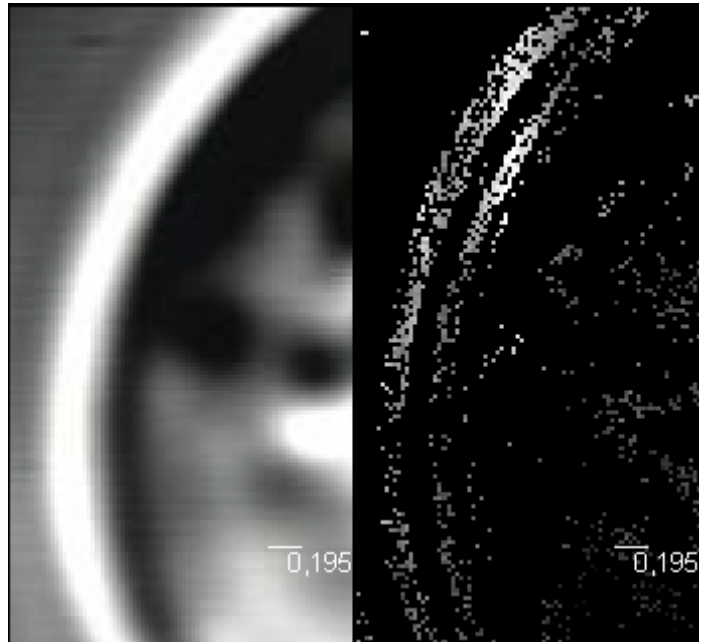

**Figure 9a. Image of cell interior showing only section of the entropy spectrum**. Image information was based on Renyi entropy contribution and calculated using $\alpha = 0,5$ White points represent points of entropy values above the treshold, black points those below the treshold. Only gray points represent resolved image information levels. From the still image the relevancy or identity of points is impossible to assess. Much better impression may be obtained from movies (see for example supplementary movie 2 and 3).

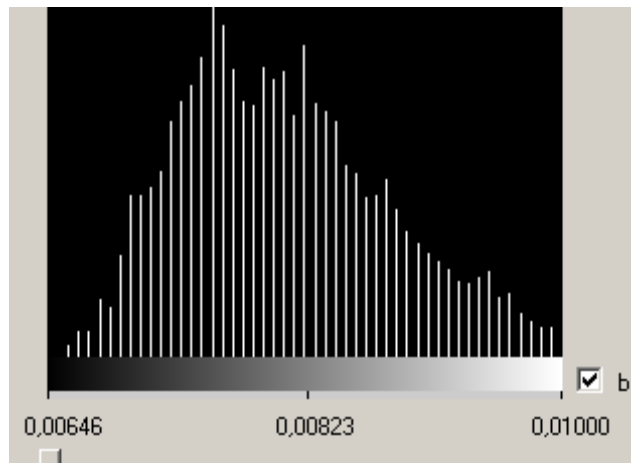

**Figure 9b. Histogram of image entropy** corresponding to the image depicted at fig. 9a.

The discussion of these results should, in my opinion, start from the following point: The videoenhanced microscopy has theoretically resolution at least 1/1000 $\lambda$ [35]. In visible light it means somewhere between 3 and 7 nm. The size of a camera point in images which are discussed here is 24.4 nm. Thus, the shape of observable objects should not be affected by the properties of the microscope optics. This should be tested experimentally. It needs to be said that trials using standard latex particles did not bring equally sharp images as we observe for living cells, namely due to very fast Brownian motion. So we have to examine the resolution using analogous electron microscopy images [39, 40]. In the article by Ossumi et al. [40], the thickness of septum formed between two dividing daughter cells of *Schizosacharomyces pombe* observed in high-pressured freezed micrograph is approximately 300 nm. The thickness of the cell wall observed in the presented sample (figure 10) is approximately 330 nm (pixel size is 24.4nm). The correspondence is almost too good if we consider structural differences which should be caused by different treatment of cells, i.e. keeping the intact native cell in our case and fast freeze of the cell at high pressure (210 Mpa). Also at the border between intracellular objects we see intensity drops of the width of 1-2 pixels. Thus, the resolution in this experiment seems to be lower than the pixel size. If this is true, the issue of discriminateability should not be risen, simply saying each different point means different information.

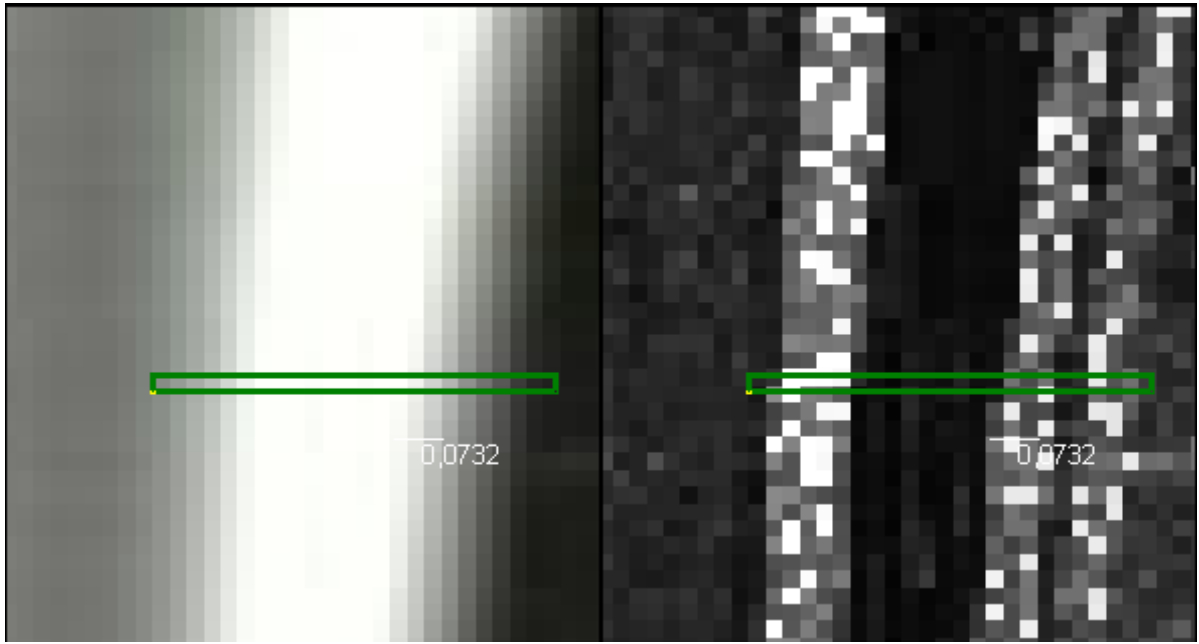

**Figure 10. Section of the image of a cell wall of S. cervisiae**. Image capture and trasformation conditions were the same at figure 9. Entropy scale expansion was applied to cover approximately 90% of all entropy values, i.e. sparse points at spectrum edges were left out. Green rectangle represents the measured object.

Observation of cells in the native state, their evolution in time and development is of course the ultimate goal in biological research. In any case, there is main difference between the videoenhanced microscopy of today and the electron microscopy, it lies in pixel size. While the scanning electron micrograph may be reproduced with in-principle infinite resolution, the camera pixel size is finite. Optical microscopy also brings a very different information, it in principle represents an optical section through the sample [41] while the electron micrograph of the freeze fractures sample represents the convex or concave surface. With the evolution of modern objectives, the thickness of the optical section decreases. Technically it is in-principle possible to build an "ideal" optical microscope in which the resolution limit is given by the density of camera points, the sensitivity is limited by the camera chip and our ability to transfer and analyse data and the structure in the z-axis may be reconstructed by the z-scan. In order to observe all details in the image, however, we have to analyse the information content of the image and reconstruct contributions of qualitatively different objects from entropy spectrum.

**In the everyday practice of biological microscopy** of present time (and, in fact any practical use of microscopy) the videoenhanced microscopy of high resolution is not used. There are basic practical problems which include mainly the necessary high light intensity which leads to sample heating and is not compatible with ordinary photographic cameras where the gain is not adjustable in broad range. In fact the development in cameras for microscopy proceeds exactly in the opposite directions, the suppliers aim to detect minute fluorescent signals from smallest diameter holes utilized to control the light path in the confocal microscope [42,43].

In biological microscopy there is even more important aspect which lies in the uniqueness of the biological sample and in the low number of proabilistically behaving sub-systems with slow time evolution (see discussions in chapter 2.2). Standard time-lapse experiment in cell biology is thus performed at ambient light intensity and using objectives with low magnification to ensure observation of large scene. Of course the ultimate answer to these problems will be technical, adaptation of videoenhanced microscopy for observation of large scenes and limiting the illumination only for the time of measurement. We follow this path together with two Czech companies, but the results are not yet ready for a report. For understanding biology we need extensive amount of biological measurements. For that we need as much as possible information from micrographs coming from everyday biological practice.

In our article [30] we report the information transformation of the time-lapse microscopy observation using standard biological setup. The cell line was grown in a growth chamber and observed by an ordinary optical microscope using phase contrast enhancement. Images were captured by ordinary photographic camera. This represents an ordinary setup utilized in medical research and results may be without obstacles utilized in practice. The first observation is that the point spread function may not be neglected. Due to the utilized phase contrast method, there are two ways how the real PSF affects the image. First is the ordinary one, the transmission of points as group of points (figure 11a). Second is the creation of hallo effects on borders of objects where there is a steep change in refractivity index (figure 11b).

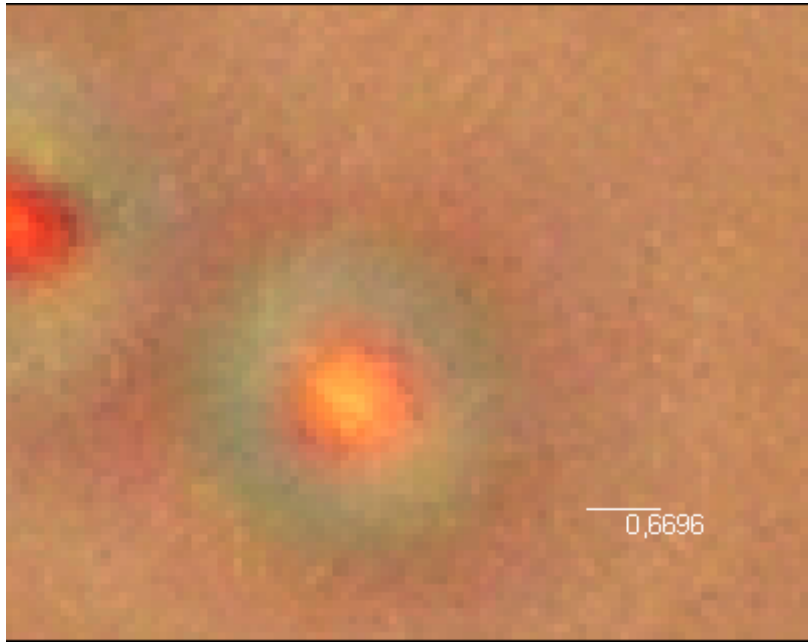

**Figure 11a Phase contrast microscope** image of latex particle of the size of 0,5 $\mu$ m.

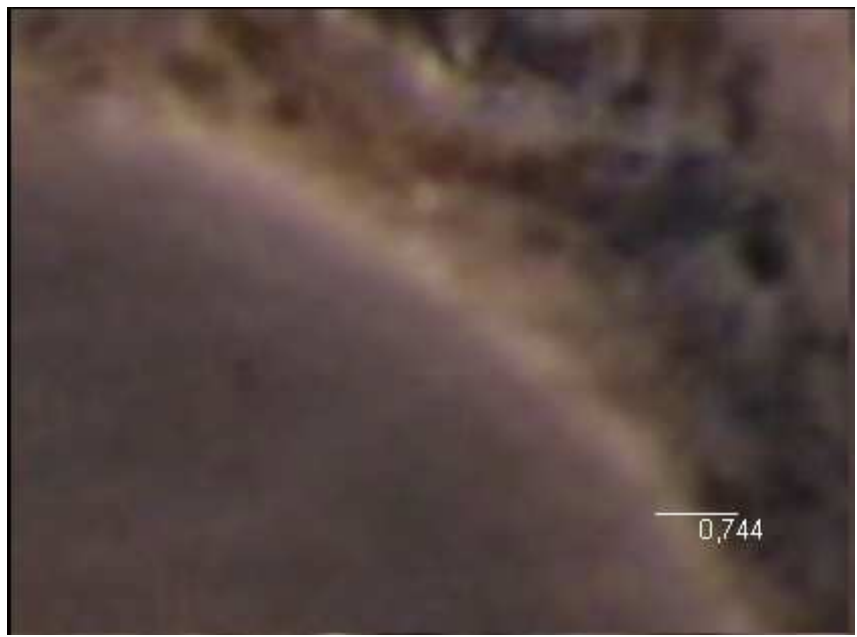

**Figure 11b Phase contrast image of HeLa cell.** Seemingly fuzzy objects are observes in the cell interior. At the cell border a large halo is formed which obscures the border position as well as object in the hallo area. The information contribution calculation (figure 12a), however, shows that specific information is still carried by each of the points.

Example of the whole image of the cell in the region comprising part of the cell nucleus is given at the figure 12a. It was created using the program *Expertomica CellMarkerSci* in which there may be analysed multiple images of the same series. In the six image windows we see the original image – central lower row, information contributions of individual channels computed as Renyi entropy contribution with coefficient $\alpha = 0{,}1$ and composite entropy image in pseudocolours (lower right image). The lower left image is image captured at pre-ceeding time instant. It is significant to see that each data point carries different information, however, the identity of this information is practically impossible to unravel.

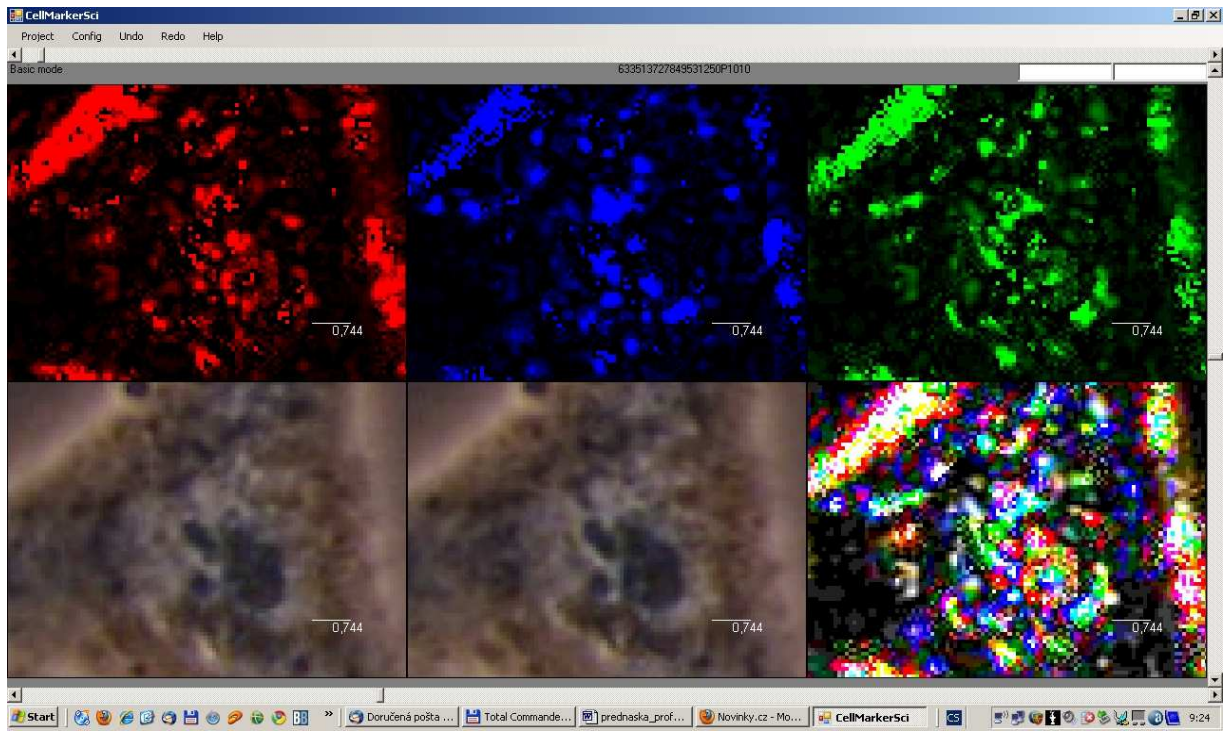

**Figure 12a** Images of original intensity and entropy distributions of a section of the image of HeLa cells. Entropy images are individual entropies at each colour channel (RGB) in the upper line, original image of examined (bottom center) and previous (bottom left) phase contrast picture and composite entropy image (bottomm right).

At figure 12b are depicted spectra of entropy distributions in composite colour channel at $\alpha$ values 0.1, 0.3, 0.5, 0.7, 1.0 and 1.5. The striking feature is the fact that, apart from $\alpha = 0.1$, we see relatively smooth intensity distributions. This indicates that different features – probabilities are spread homogeneously over the spectrum. This is in contrast to analysis of the videoenhanced microscopy spectrum. Perhaps we may speculate that this is caused by the fact the imaging process is in the case of "standard" microscopy with relatively large depth of focus is dominated by the shape of the point spread function. And this evolves smoothly through the depth of focus. The observed distributions then reflects the position of the object relatively to the focus rather then properties of objects themselves.

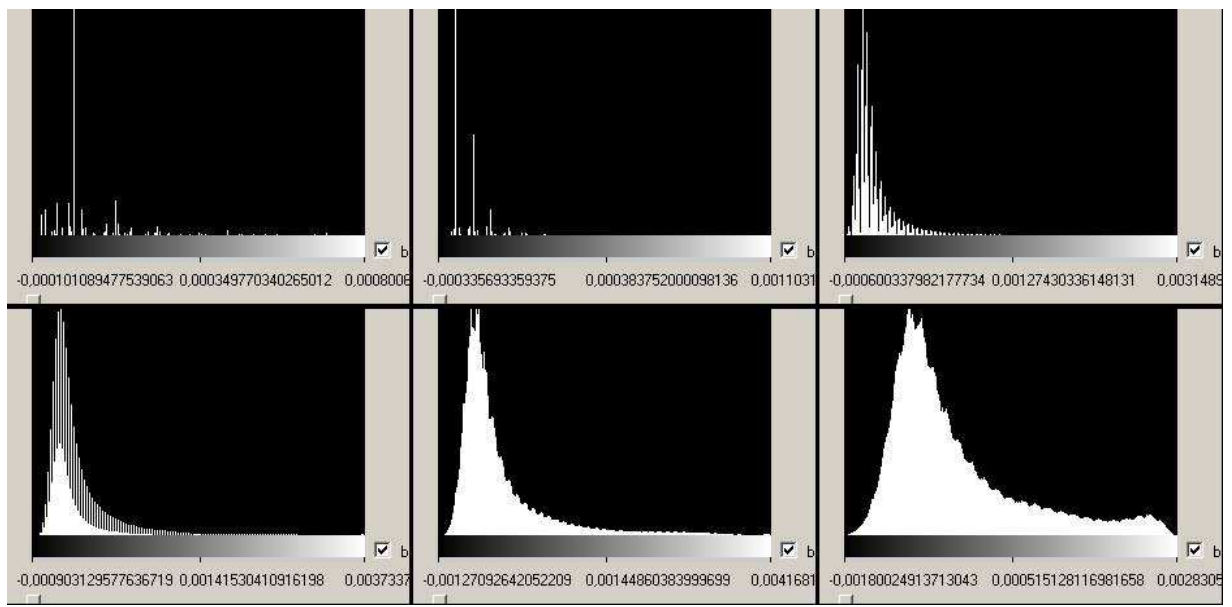

**Figure 12b Entropy spectra of HeLa cells images in RGB space** using $\alpha = 0.1, 0.3, 0.5, 0.7, 1.0, 1.5$. Spectral lines are projection of apparent 3D – wavelength – intensity – entropy spectrum to the intensity – entropy plane.

The scene in this image is still insufficient to obtain reasonable statistics of cell division, as was discussed in the chapter 2.2, but the number of observed individual dynamic objects inside cells is significant. This experiment may be used as basis for construction of biological model.

## 2.4 Model of cell monolayer as stochastic system – state of the art

The study described in the chapter 2.3 was initiated by the request of biomedical practicioners to identify and classify cells in cell monolayer. If we want to follow the model construction which was proposed in the chapter 2.2, automated identification of cells as expected elements of the system is primary. This task did not show up to be as easy as the manuals for the cell recognition software packages would suggest. The proper approach turned to be distinction between the recurrently moving objects – i.e. objects wobbling in approximately the same position for long periods of time – and stable objects – the background (supplementary video 2 and 3). This, and many other, observations [30] brought about the overall approach to sub-system identification in living cells: We should seek objects which recurrently return to approximately the same position in space and in coloration. The experiments described in part 2.3 give us almost incomprehensible amount of information to analyse.

At videos 4-6 is shown original cell monolayer scene at the beginning of the experiment. On videos 7-9 are shown recurrent movements inside individual cell and on videos 10-12 is depicted cell-to-cell communication. Similar videa may be prepared from various parts of the cell and at various times of cell development. A serious trouble in interpretation lies in the absence of comparable images. There are two future pathways which we follow (1) automated identification and classification of all recurrently behaving objects in the cell and (2) identification of individual objects with known cell compartments for which are available specific coloration agents.

The ultimate goal should, however, be description of cell state by a set of phenomenological variables. We have to ask what are the variables $V_r$ which we should measure to be able to estimate the causal relations $\Xi_r$ and probabilities. There is at the moment no comprehensive list, but based on observations like that shown at the supplementary video 18 we assume that the available phenomenological variables will be shape dynamics, colour dynamics and frequency. We may denote the phenomenological sub-system as

$$\pi_{r,\phi} = (T, V_{r,\phi}, \Xi_{r,\phi}, P_{r,\phi})  \tag{14}$$

Where $V_{r,\phi}, \Xi_{r,\phi}, P_{r,\phi}$ are observable variables, observable causal relations and observable probabilities respectively. In this way, we arrive to the best available definition of living cells monolayer.

# 3 Conclusions

Videorecords of living cells were always welcome with great enthusiasm. The state of the art may be documented on the article of Neumann et al. [44]. The main message is that the contemporary movies focus on fluorescence labelled cells which simultaneously identify maximally three molecules. As we argued earlier [29], in order to describe the state of the cell, we need to define and determine the equivalent of the Gibbs energy for the particular cell. The fluorescent markers are equivalent to determination of a concentration and that only if the marking procedure is focused on a protein or metabolite molecule.

In equilibrium case for the Gibbs energy $G$ holds

$$G = G_0 + F(c_1, c_2, \ldots c_n, p, V, T)  \tag{14}$$

where $F(c_1, c_2, \ldots c_n, p, V, T)$ is in fact a version of the system state function of the system which includes all molecular species in the system $c_1, c_2, \ldots c_n$ and other relevant variables such as pressure, volume or temperature: $p, V, T$. Certainly some concentrations dominate the system but even in inorganic mixtures the deviation from ideality (i.e. from the ideal gas state function) are in he order of 40%.

In assessment of the living system state there are utilized several approaches. The genomic approach is a static one, it examines the genome, the DNA contents, which is identical for all the cells. For the purpose of building the state model it represents examination of the potential for cell state. The transcriptomic approach mapping the m-RNA which codes for concrete proteins, which was also utilized in [45], indicates which proteins may be synthesizes in a particular, concrete, cell. The proteomic and metabolomic approach examines the composition of molecules which finally participate in the intramolecular chemistry, proteins and metabolites. But these molecules are practically unavailable for measurement inside the living cell.

The approach which we follow suggest completely different approach, direct measurement of the Gibbs energy equivalent rather than deducing it from individual concentrations. The recipe what is we should seek is given in [25]. It says that we should seek probabilities of individual trajectories $\mu$ in a dynamic system whose state space is separated in so-called generating Markov partition. This means that the probability of transitions between partitions is independent from the history. Which is exactly the same requirement as it was set up by Žampa [20-23] in the general stochastic system theory. There it is required to examine the history of the system from the observed state transition up to the point when we can determine unique trajectory element $(C_{k,l}, D_{k,l})$. To each of these trajectory elements we may assign individual probability density function $p_{k,l}$. Žampa demonstrates that such definition of state of a dynamic system is the only one which may be identified with common understanding of state in mechanics or thermodynamics. Thus, to measure system state by the approach proposed in the equation (14) is fully conform with the non-linear dynamic analysis. The trajectory probability $\mu_s$ would than be

$$\mu_s = f(p_{k,l,s,q}) \tag{15}$$

the functional of all elementary probability density functions $p_{k,l,s,q}$ for all elements $q$ of a given trajectory $s$. It holds that $\sum_1^{s_{max}} \mu_s = 1$ for all available trajectories. Vattay in [24] proposes that the state – or partition sum should be given by

$$Z(\beta) = \sum_1^n \mu_s^\beta \tag{15}$$

Where n is the number of available trajectories in the system and $\beta$ is equivalent to $\dfrac{1}{T}$ in thermodynamics, a global parameter representing the state of the system. For all trajectories of the system and the free energy analogy would be the Rényi entropy

$$H_\beta = \lim_{n \to \infty} \frac{1}{n} \frac{1}{1-\beta} \log_2 \sum_1^n \mu_s \tag{16}$$

In this text I report the technical, methodological and theoretical setup for analysis of development of a monolayer of living cells as non-linear dynamic system. I also propose how the state of the system should be determined and identified. The proposed method is correct from the point of view of both non-linear dynamics and stochastic systems theory and outline the sources of this identity. Numerous concrete experiments have to be performed.

However, question should be asked whether we shall ever be able to propose to a medical doctor a $H_\beta$ value to discriminate between two cancer diagnoses. At least, we have the frame for evaluation. Certainly we are far from the $n \to \infty$ limit. To approximate the state, we may, as in the case of image analysis and in fact by the same method we may calculate experimental probabilities and from them the experimental Rényi entropy

$$H_{\beta,\exp} = \frac{1}{n} \frac{1}{1-\beta} \log_2 \sum_1^n \mu_s \tag{17}$$

for $n$ observable trajectories. In the experimental setup in which we shall be able to change experimental conditions, i.e. by addition of potential drug substance, we shall obtain a new set of trajectories. Eventually we may obtain the equivalent of the potential gradient as

$$\left(\frac{\partial H_\beta}{\partial c_i}\right)_{c_{j \neq i}, T, V, p} \approx \frac{\Delta\left(\dfrac{1}{1-\beta} \log_2 \sum_1^n \mu_s\right)}{\Delta c_i} \tag{17}$$

as an approximation of the response of a particular cancer tissue to a given chemotherapeutics. All technical conditions for such measurement are now in our hands.

## Acknowledgements

I am indebted mainly to members of my team and students.

To the work presented in this paper contributed following researchers: V. Brezina, S. Kucerova and the team of the cell culture laboratory contributed by cell cultivation.

Images of latex particles utilized in point spread function estimation were performed by P. Císař and T. Náhlík.

The author of the concept of image entropy calculation if J. Urban and it was further developed together with J. Vaněk and P. Císař which resulted in *Expertomica Entropy calculator* software. The *Expertomica CellMarker* software was developed by P. Císař as well as few other tools utilized in this work. Authors of the *Expertomica Cells* software are Tomáš Levitner and Štěpán Timr.

G. v Buquoys manuscripts were brought to me by M. Vlačihová who also participated on preparation of respective manuscripts

Videoenhanced microscopy images were prepared by prof. Irene Leichtscheidel from the Vienna university.

H. Duadi, O. Margalit, Z. Zalevski from *Bar-Ilan University and* V Sarafis from University of Melbourne and University of South Bohemia provided us with manuscript of their paper [41]

Finally, this work would not be possible without discussions and encouragement of P. Jizba from the Faculty of Nuclear Sciences and Physical Engineering, Czech Technical University

I woudl also like to thank all the others who helped me on my way to science. Namely to my family which, as a major surprise, is almost as enthusiastic in the goals which I try to achieve as myself. This ceratinly is a miracle.

# 4 Appendix 1: description of supplementary material

Supplementary video 1 – Hela cells development

In the video is shown the most typical experiment in present day cell biology. It shows development of a monolayer of He-La cell in time lapse. The time difference between individual frames was one minute. Altogether 5000 frames were captured. From these data images and graphs shown at figures 5 and 6 were constructed.

Supplementary video 2 – Development of section of the image entropy spectrum in time.

Timeframes were captured at 1/25s intervals and are displayed at ½ s intervals. In similar way al sections of the entropy spectrum are now examined to find out what is the origin of their appearance.

Supplementary video 3-4 – Cell border identification.

Detection of cell borders is an extremely difficult task in standard still phase contrast images. Using time-lapse movies we developed a prototype software which detects groups of objects showing recurrent behavior in time, position and spectral characteristics. These are cells or their groups. In the short future this approach will be extended to include identification of individual cells in cell layers and identification of recurrent intracellular objects.

Supplementary videa 4-6 time development of original image scene of cell monolayer using entropies with $\alpha = 0.1, 0.7, 1.5$

Supplementary videa 7-9 show development of recurrent objects inside a cell using entropies with $\alpha = 0.1, 0.7, 1.5$

Supplementary videa 10-12 show cell-to-cell communication via mass transfer (communication bond) using entropies with $\alpha = 0.1, 0.7, 1.5$

# 5 Appendix 2: references to used methods

Methods of cultivation of HeLa cells and image capture are described in paper [28]
Yeast cells were cultivated in the *Saccharomyces cerevisiae* rich (complex) medium YPD (1l) which consoisted of 20g Bacto peptone, 10g Yeast Extract, 950 ml of water with pH 5,9 adjusted by with HCl

Videoenhanced microscopy was performed as described in [29]

Image transformations were performed as described in the manuscript [29]

Video transformations were performed in the freeware VirtualDub (www.vistualdub.org)

# 6 Appendix 3 Example of use of the method in material enginnering

Methods described in this article are of general interest in any field where microscopy techniques are used. Recently we have solved the problem of recognition of polymer polycrystalline objects for quality control in car industry. Images are captured by polarization microscopy. The quality control is performed by manual inspection of images which is subjective and overlooks a lot of details. The entropy calculation method which transforms images and is able to emphasise numerous features opens the possibility to combine series of images for recognition. At images A3-1, A3-2 and A3-3 are shown data which illustrate the method.

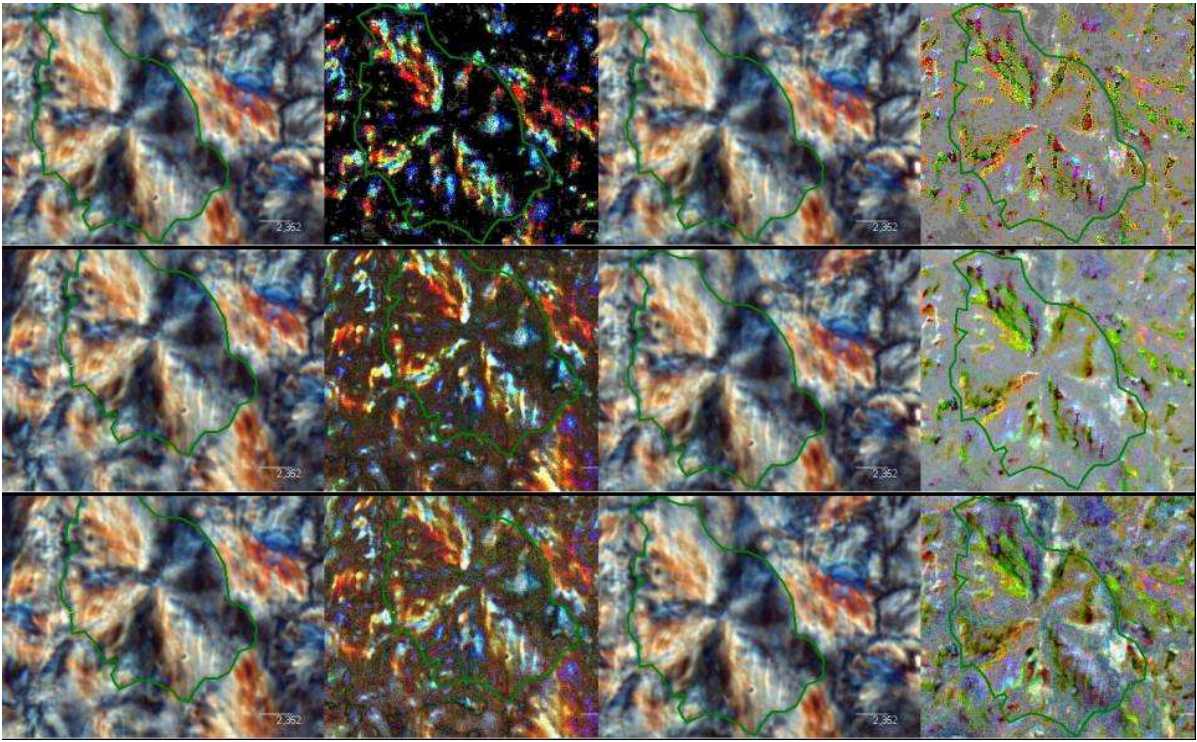

**Figure A3-1** Entropy transformed images of one polycrystalline nucleus. Left column from up to down are entropy transformations of the image into image entropy space using